\documentclass[a4paper]{llncs}

\usepackage[utf8]{inputenc}
\usepackage[T1]{fontenc}
\usepackage[english]{babel}
\usepackage{microtype}
\usepackage{graphicx}
\usepackage{url}
\usepackage{complexity}

\usepackage{psystems}

\newcommand{\set}[1]{\{#1\}}

\newcommand{\yes}{\mathsf{yes}}
\newcommand{\no}{\mathsf{no}}
\newcommand{\pos}[2][]{\varphi^{#1}(#2)}

\title{A Turing Machine Simulation\\by P~Systems without Charges}
\author{
  Alberto Leporati\inst{1} \and
  Luca Manzoni\inst{1} \and
  Giancarlo Mauri\inst{1} \and\\
  Antonio E. Porreca\inst{1,2} \and
  Claudio Zandron\inst{1}
}

\institute{
  Dipartimento di Informatica, Sistemistica e Comunicazione\\
  Università degli Studi di Milano-Bicocca\\
  Viale Sarca 336, 20126 Milano, Italy\\
  \email{\{leporati,luca.manzoni,mauri,zandron\}@disco.unimib.it}
  \and
  Aix Marseille Université, Université de Toulon, CNRS, LIS, Marseille, France
  \email{antonio.porreca@lis-lab.fr}
}

\begin{document}

\maketitle

\begin{abstract}
It is well known that the kind of P~systems involved in the definition of the P conjecture is able to solve problems in the complexity class $\P$ by leveraging the uniformity condition. Here we show that these systems are indeed able to simulate deterministic Turing machines working in polynomial time with a weaker uniformity condition and using only one level of membrane nesting. This allows us to embed this construction into more complex membrane structures, possibly showing that constructions similar to the one performed in~\cite{Leporati2014e} for P~systems with charges can be carried out also in this case.
\end{abstract}

\section{Introduction}
\label{sec:introduction}

The construction of P~systems simulating Turing machines (TM) using as few membranes (or cells) as possible and limiting the depth of the system is one of the ``tricks'' that allowed the nesting of multiple machines to solve problems in large complexity classes. For example, nesting of non-deterministic machines (where the non-determinism was simulated by membrane division) and a counting mechanism allows to characterize $\P^{\#\P}$, the class of all problems solvable by a deterministic TM with access to a $\#\P$ oracle~\cite{Leporati2014e,Leporati2017a}. The same ideas can be applied to tissue P~systems~\cite{MartinVide2003a}, where the different communication topology makes even more important to keep TM simulations compact~\cite{Leporati2017e}.

The P conjecture is a long-standing open problem in membrane computing first presented in 2005~\cite[Problem F]{Paun2005a} that, in its essence, asks what is the power of one charge when compared to two charges. We feel that one important step to determine the computational power of active membrane systems without charges and with membrane dissolution is to see which is the minimal system able to simulate a deterministic polynomial-time TM. Here we show that a shallow system is sufficient to perform such a simulation \emph{without} delegating everything to the machine of the uniformity condition. Hopefully, this construction will allow us to define systems in which different TM can be ``embedded'' at different levels in a large membrane structure, thus making possible to mimic the construction performed in~\cite{Leporati2014e} for P~systems with charges.

This paper is organized as follows: Section~\ref{sec:basic-notions} will recall some basic notions on P~systems. The main construction and result is presented in Section~\ref{sec:tm-simulation}, while ideas for further research are presented in Section~\ref{sec:conclusions}.

\section{Basic Notions}
\label{sec:basic-notions}

For an introduction to membrane computing and the related notions of formal language theory and multiset processing, we refer the reader to \emph{The Oxford Handbook of Membrane Computing}~\cite{Paun2010a}. Here we recall the formal definition of P~systems with active membranes using weak non-elementary division rules~\cite{Paun2001a,Zandron2008a}.

\begin{definition}
  A \emph{P~system with active membranes with dissolution rules} of initial degree~$d \ge 1$ is a tuple
  \begin{align*}
    \Pi = (\Gamma, \Lambda, \mu, w_{h_1}, \ldots, w_{h_d}, R)
  \end{align*}
  where:
  \begin{itemize}
    \item $\Gamma$ is an alphabet, i.e., a finite non-empty set of symbols, usually called \emph{objects};
    \item $\Lambda$ is a finite set of labels;
    \item $\mu$ is a membrane structure (i.e., a rooted \emph{unordered} tree, usually represented by nested brackets) consisting of~$d$ membranes labelled by elements of~$\Lambda$ in a one-to-one way;
    \item $w_{h_1}, \ldots, w_{h_d}$, with~$h_1, \ldots, h_d \in \Lambda$, are multisets (finite sets with multiplicity) of objects in~$\Gamma$, describing the initial contents of each of the~$d$ regions of~$\mu$;
    \item $R$ is a finite set of rules.
  \end{itemize}
\end{definition}

\noindent
The rules in~$R$ are of the following types:
\begin{enumerate}
  \item[(a)] \emph{Object evolution rules}, of the form~$\pevolvez{h}{a}{w}$. \\
  They can be applied inside a membrane labelled by~$h$ and containing an occurrence of the object~$a$; the object~$a$ is rewritten into the multiset~$w$ (i.e.,~$a$ is removed from the multiset in~$h$ and replaced by the objects in~$w$).

  \item[(b)] \emph{Send-in communication rules}, of the form $\psendinz{h}{a}{b}$. \\
  They can be applied to a membrane labelled by $h$ and such that the external region contains an occurrence of the object $a$; the object $a$ is sent into $h$ becoming $b$.

  \item[(c)] \emph{Send-out communication rules}, of the form~$\psendoutz{h}{a}{b}$. \\
  They can be applied to a membrane labelled by~$h$ and containing an occurrence of the object~$a$; the object~$a$ is sent out from~$h$ to the outer region becoming~$b$.

  \item[(d)] \emph{Dissolution rules}, of the form~$\pdissolvez{h}{a}{b}$.\\
  They can be applied to a non-skin  membrane labelled by~$h$ and containing an occurrence of the object~$a$; the object~$a$ is sent out from~$h$ to the outer region becoming~$b$, the membrane~$h$ ceases to exist and all the other objects it contains are sent into the outer region.
\end{enumerate}

A computation step changes the current configuration according to the following principles:
\begin{itemize}
  \item The application of rules is \emph{maximally parallel}: each object appearing on the left-hand side of evolution, communication, or division rules must be subject to exactly one of them. Analogously, each membrane can only be subject to one communication or dissolution rule (types (b)--(d)) per computation step; for this reason, these rules will be called \emph{blocking rules} in the rest of the paper. As a result, the only objects and membranes that do not evolve are those associated with no rule.
  \item When several conflicting rules can be applied at the same time, a nondeterministic choice is performed; this implies that, in general, multiple possible configurations can be reached after a computation step.
  \item In each computation step, all the chosen rules are applied simultaneously in an atomic way. However, in order to clarify the operational semantics, each computation step is conventionally described as a sequence of micro-steps whereby each membrane evolves only after its internal configuration (including, recursively, the configurations of the membrane substructures it contains) has been updated.
  \item The outermost membrane (the root of the membrane structure) cannot be divided, and any object sent out from it cannot re-enter the system again.
\end{itemize}
A \emph{halting computation} of the P~system~$\Pi$ is a finite sequence~$\compC = (\confC_0, \ldots, \confC_k)$ of configurations, where~$\confC_0$ is the initial configuration, every~$\confC_{i+1}$ is reachable from~$\confC_i$ via a single computation step, and no rules of~$\Pi$ are applicable in~$\confC_k$.

P~systems can be used as language \emph{recognisers} by employing two distinguished objects~$\yes$ and~$\no$: we assume that all computations are halting, and that either one copy of object~$\yes$ or one of object~$\no$ is sent out from the outermost membrane, and only in the last computation step, in order to signal acceptance or rejection, respectively. If all computations starting from the same initial configuration are accepting, or all are rejecting, the P~system is said to be \emph{confluent}.

In order to solve decision problems (or, equivalently, decide languages), we use \emph{families} of recogniser P~systems $\familyPi = \{ \Pi_x : x \in \Sigma^\star \}$. Each input~$x$ is associated with a P~system~$\Pi_x$ deciding the membership of~$x$ in a language $L \subseteq \Sigma^\star$ by accepting or rejecting. The mapping~$x \mapsto \Pi_x$ must be efficiently computable for inputs of any length, as discussed in detail in~\cite{Murphy2011a}.
\begin{definition}
  \label{def:uniform}
  A family of P~systems~$\familyPi = \{ \Pi_x : x \in \Sigma^\star \}$ is \emph{(polynomial-time) uniform} if the mapping~$x \mapsto \Pi_x$ can be computed by two polynomial-time deterministic Turing machines~$E$ and~$F$ as follows:
  \begin{itemize}
    \item $F(1^n) = \Pi_n$, where~$n$ is the length of the input~$x$ and~$\Pi_n$ is a common P~system for all inputs of length~$n$, with a distinguished input membrane.
    \item $E(x) = w_x$, where~$w_x$ is a multiset encoding the specific input~$x$.
    \item Finally,~$\Pi_x$ is simply~$\Pi_n$ with~$w_x$ added to a specific membrane, called the  \emph{input membrane}.
  \end{itemize}
  The family~$\familyPi$ is said to be (polynomial-time) semi-uniform if there exists a single deterministic polynomial-time Turing machine~$H$ such that~$H(x) = \Pi_x$ for each~$x \in \Sigma^\star$.
\end{definition}

Any explicit encoding of~$\Pi_x$ is allowed as output of the construction, as long as the number of membranes and objects represented by it does not exceed the length of the whole description, and the rules are listed one by one. This restriction is enforced in order to mimic a (hypothetical) realistic process of construction of the P~systems, where membranes and objects are presumably placed in a constant amount during each construction step, and require actual physical space proportional to their number; see also~\cite{Murphy2011a} for further details on the encoding of P~systems.

\section{Simulation of Polynomial-time Turing machines}
\label{sec:tm-simulation}

In this section we provide a simulation of a deterministic TM working in polynomial time by a P system that uses only one level of nesting. Any information exchange between objects can happen only via dissolution. By applying different evolution rules, it is possible for an object to detect whether it is inside or outside an elementary membrane (i.e., to ``know'' if the elementary membrane where it was has been dissolved). By combining this mechanism with a timer, it is also possible to encode the time when the membrane was dissolved, thus allowing to evolve in different ways according to this additional information.

Let $M$ be a polynomial-time deterministic TM having alphabet $\Sigma$, set of states $Q$, and transition function $\delta: Q \times \Sigma \to Q \times \Sigma \times \set{-1,+1}$. We assume that, for an input of length $n$ machine $M$ halts in time $p(n)$ and, thus, it uses no more than $p(n) + 1$ cells. We are going to define a P~system $\Pi$ that simulates the computation of $M$ in $O(p(n)|\Sigma|)$ steps. That is, the simulation of every step of $M$ will require a number of steps in $\Pi$ that is proportional to the size of the alphabet of $M$, thus providing an efficient simulation.

The P~system $\Pi$ has $(p(n)+1)^2 + p(n)^2 + p(n) + 1$ labels, one for the skin membrane and two for each pair of time and position in the TM tape:
\begin{align*}
  \Lambda = & \set{0} \cup \set{(i,j) \;|\; i,j \in \set{0,\ldots,p(n)}}\\
  & \cup \set{(i,j)' \;|\; i \in \set{0,\ldots,p(n)},\; j \in \set{0, \ldots, p(n) - 1}} \quad.
\end{align*}
Since we assume that no kind of membrane division is present, in the following we can identify membranes with labels, since each label is used by exactly one membrane. The semantics of the labels is that a  membrane with label $(i,j)$ will represent the $i$-th cell of the TM tape at time $j$. The additional membrane $(i,j)'$ is used in performing the transition between time steps $j$ and $j+1$, which also explains why the label is not present for time $p(n)$.

The set of objects of the simulating P~system will be:
\begin{align*}
  \Gamma = & \set{a_{i,j,k} \;|\; i,j \in \set{0,\ldots,p(n)}, \; 0 \le k < m + 5, a \in \Sigma} \\
           & \cup \set{q_{i,j,k} \;|\; i,j \in \set{0,\ldots,p(n)}, \; 0 \le k \le m + 5, q \in Q} \\
           & \cup \set{q_{i,j,k,a} \;|\; i,j \in \set{0,\ldots,p(n)}, \; 0 \le k \le m + 5, \; q \in Q,\; a \in \Sigma} \\
           & \cup \set{a_i \;|\; a \in \Sigma, i \in \set{0,\ldots,p(n)}} \cup \set{q^I}
\end{align*}
where $m = |\Sigma|$ and $q^I$ is the initial state of the TM. The first three sets of the union represent, respectively, the symbols on the tape, the states of the TM, and the states of the TM together with the symbol currently present under the tape head. The last two sets are only used to encode the initial configuration of the TM. The value of $k$ ranges from $0$ to $m + 5$ because each step of the TM will be simulated in $m + 5$ time steps.

\begin{figure}
  \centering
  \includegraphics[width=0.48\textwidth,page=1,trim=5cm 6cm 5cm 6cm,clip]{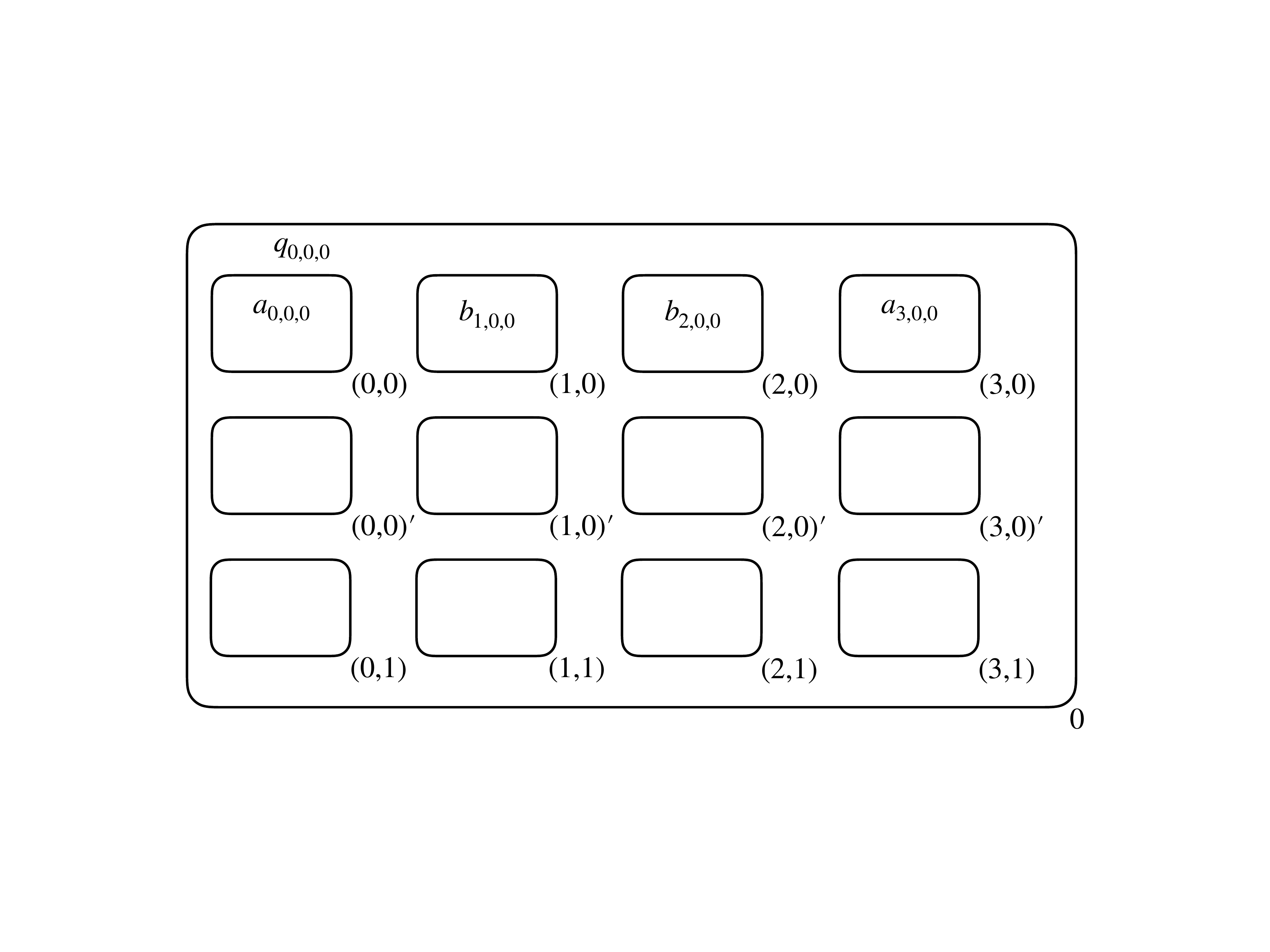}
  \includegraphics[width=0.48\textwidth,page=2,trim=5cm 6cm 5cm 6cm,clip]{Simulation.pdf}\\
  \hfill $t=1$ \hfill \hfill $t=2$ \hfill\mbox{}\vspace{0.5em} \\
  \includegraphics[width=0.48\textwidth,page=3,trim=5cm 6cm 5cm 6cm,clip]{Simulation.pdf}
  \includegraphics[width=0.48\textwidth,page=4,trim=5cm 6cm 5cm 6cm,clip]{Simulation.pdf}\\
  \hfill $t=3$ \hfill \hfill $t=4$ \hfill\mbox{}\vspace{0.5em} \\
  \includegraphics[width=0.48\textwidth,page=5,trim=5cm 6cm 5cm 6cm,clip]{Simulation.pdf}
  \includegraphics[width=0.48\textwidth,page=6,trim=5cm 6cm 5cm 6cm,clip]{Simulation.pdf}\\
  \hfill $t=5$ \hfill \hfill $t=6$ \hfill\mbox{}\vspace{0.5em} \\
  \includegraphics[width=0.48\textwidth,page=7,trim=5cm 6cm 5cm 6cm,clip]{Simulation.pdf}
  \includegraphics[width=0.48\textwidth,page=8,trim=5cm 6cm 5cm 6cm,clip]{Simulation.pdf}\\
  \hfill $t=7$ \hfill \hfill $t=8$ \hfill\mbox{}\vspace{0.5em} \\
  \includegraphics[width=0.48\textwidth,page=9,trim=5cm 6cm 5cm 6cm,clip]{Simulation.pdf}
  \includegraphics[width=0.48\textwidth,page=10,trim=5cm 6cm 5cm 6cm,clip]{Simulation.pdf}\\
  \hfill $t=9$ \hfill \hfill $t=10$ \hfill\mbox{}\vspace{0.5em} \\
  \caption{\label{fig:one-step} The simulation of one computation step of the TM $M$ by means of a P~system $\Pi$. The alphabet $\Sigma$ is $\set{a,b}$ and the tape contains four cells.}
\end{figure}

Let $a_1,a_2, \ldots, a_{p(n)}$ be the initial contents of the TM tape. It is encoded in the initial configuration of $\Pi$ as the objects $a_{1,1},a_{2,2}, \ldots, a_{p(n),p(n)}$ inside the skin membrane. As an example, if the initial content of the tape is $abba$, then it will be encoded by the multiset $a_1b_2b_3a_4$. The initial state $q^I$ is encoded by the object $q^I$. The following rules send the objects representing the TM tape inside the corresponding membranes: the object $a_i$ is sent into the membrane $(i, 0)$ and is rewritten as $a_{i,0,0}$. At the same time the object $q^I$ is rewritten as $q^I_{0,0,0}$:
\begin{align*}
  & \psendinz{(i,0)}{a_i}{a_{i,0,0}} & \text{for $a \in \Sigma$} \\
  & \pevolvez{0}{q^I}{q^I_{0,0,0}}
\end{align*}
These rules will not be further applied during the simulation. After this first ``bookkeeping'' step the actual simulation of one TM step can start; see Fig.~\ref{fig:one-step} for an example.

Let $\varphi$ be a bijection from $\Sigma$ to $\set{1, \ldots, m}$ providing a total ordering of the TM alphabet. The main idea is to have each object representing the symbol $a$ written on position $i$ at time $j$ on the TM tape dissolving the membrane $(i,j)$ when its subscript is $i, j, \pos{a}$. This means that any other object present in the same membrane (in our case, the object representing the current state of the TM) can infer the symbol under the tape head and act accordingly. The evolution of the objects representing the tape content for the first $m+1$ time steps of each TM step simulation is described by the following rules:
\begin{align*}
  & \pevolvez{(i,j)}{a_{i,j,k}}{a_{i,j,k+1}} & \text{for $0 \le k < \pos{a}$ and $a \in \Sigma$} \\
  & \pdissolvez{(i,j)}{a_{i,j,k}}{a_{i,j,k+1}} & \text{for $k = \pos{a}$ and $a \in \Sigma$} \\
  & \pevolvez{0}{a_{i,j,k}}{a_{i,j,k+1}} & \text{for $\pos{a} < k \le m$ and $a \in \Sigma$}
\end{align*}
Notice how the objects simply ``count'' in the subscript, except that when $k = \pos{a}$ the membrane in which they are contained is dissolved.

At the same time the object representing the TM state enters the membrane $(i,j)$, representing that the tape head at time $j$ is in position $i$ and starts to count. When membrane $(i,j)$ is dissolved it is possible to infer the object that dissolved it, and thus the symbol on the tape under the tape head, which is represented by $\pos[-1]{a}$ (which is well defined since $\varphi$ is a bijection between $\Sigma$ and $\set{1,\ldots,m}$. The corresponding rules are:
\begin{align*}
  & \psendinz{(i,j)}{q_{i,j,0}}{q_{i,j,1}} & \text{for $q \in Q$} \\
  & \pevolvez{(i,j)}{q_{i,j,k}}{q_{i,j,k+1}} & \text{for $1 \le k \le m$ and $q \in Q$} \\
  & \pevolvez{0}{q_{i,j,k}}{q_{i,j,k+1,\pos[-1]{k}}} & \text{for $1 \le k \le m$, and $q \in Q$}\\
  & \pevolvez{0}{q_{i,j,k,a}}{q_{i,j,k+1,a}} & \text{for $1 \le k \le m$, $a \in \Sigma$, and $q \in Q$}
\end{align*}
At time step $m+1$ in the simulation of the current TM step, all membranes with label $(i,j)$ (for all $i$ and with $j$ the current TM step being simulated) have been dissolved. Now the object representing the TM state continues to wait in the skin membrane while \emph{all} the objects representing the TM tape are sent in into the corresponding membranes $(i,j)'$. These membranes will be employed to delete the current content of the cell under the TM head and to replace it with the new symbol. The rules applied at time step $m+1$ are the following ones:
\begin{align*}
  & \psendinz{(i,j)'}{a_{i,j,m+1}}{a_{i,j,m+2}} & \text{for $a \in \Sigma$} \\
  & \pevolvez{0}{q_{i,j,m+1,a}}{q_{i,j,m+2,a}} & \text{for $q \in Q$ and for $a \in \Sigma$}
\end{align*}
Ones all the objects of the form $a_{i,j,k}$ have entered the membranes $(i,j)'$, they wait for the object representing the TM state to enter:
\begin{align*}
  & \pevolvez{(i,j)'}{a_{i,j,m+2}}{a_{i,j,m+3}} & \text{for $a \in \Sigma$} \\
  & \psendinz{(i,j)'}{q_{i,j,m+2,a}}{q_{i,j,m+3,a}} & \text{for $q \in Q$ and $a \in \Sigma$}
\end{align*}
At time step $m+3$ the membrane containing the object representing the TM state is dissolved. In all other membranes the objects representing the TM tape wait for one more step:
\begin{align*}
  & \pevolvez{(i,j)'}{a_{i,j,m+3}}{a_{i,j,m+4}} & \text{for $a \in \Sigma$} \\
  & \pdissolvez{(i,j)'}{q_{i,j,m+3,a}}{q_{i,j,m+4,a}} & \text{for $q \in Q$ and $a \in \Sigma$}
\end{align*}
One of the focal point of this simulation algorithm happens at time step $m+4$ (always relative to the start of the simulation of the current TM step). Here, all the objects representing the tape content dissolve the membrane $(i,j)'$ in which they are in. The \emph{only} object not performing this step is the one that was sent into the skin membrane by the dissolution triggered by the object representing the TM state. That object is deleted (by being rewritten into the empty multiset $\epsilon$) and the state object produces its replacement according to the transition function $\delta$ of the TM:
\begin{align*}
  & \pdissolvez{(i,j)'}{a_{i,j,m+4}}{a_{i,j,m+5}} & \text{for $a \in \Sigma$} \\
  & \pevolvez{0}{a_{i,j,m+4}}{\epsilon} & \text{for $a \in \Sigma$} \\
  & \pevolvez{0}{q_{i,j,m+4,a}}{q_{i,j,m+5,a} b_{i+d,j,m+5}} & \text{for $q \in Q$,  $a \in \Sigma$,}\\
  & & \text{and $\delta(q,a) = (r,b,i+d)$}
\end{align*}
Notice that the state object will be actually rewritten from $q$ to $r$ during the next time step. Finally, the simulation of the next TM step can start by sending in all the objects representing the TM tape to the membranes $(i,j+1)$ and resetting the last component of their subscript. At the same time the object representing the TM state actually applies the transition function $\delta$ and rewrites itself:
\begin{align*}
  & \psendinz{(i,j+1)}{a_{i,j,m+5}}{a_{i,j+1,0}} & \text{for $a \in \Sigma$} \\
  & \pevolvez{0}{q_{i,j,m+5,a}}{r_{i+d,j+1,0}} & \text{for $q \in Q$, $a \in \Sigma$,} \\
  & & \text{and $\delta(q,a) = (r,b,i+d)$}
\end{align*}

Notice that all rules, labels, and objects can be constructed by a logarithmic space TM. In fact, most of them are constructed by iterating either a constant or a polynomial number of times to produce the necessary subscripts. Since the counters are all at most polynomial in the number that they contain, they can be encoded in a logarithmic number of bits.

We can thus state the main result:
\begin{theorem}
  $(\L,\L)$-uniform families of confluent \emph{shallow} P~systems with active membranes with dissolution and without division can solve all problems in $\P$.
\end{theorem}
The result was already known for non-shallow system~\cite{Murphy2011a} but here there are two main innovations: the systems here are \emph{shallow}, i.e., of depth~$1$, and the construction is via a direct simulation of a Turing machine, which allows one to embed this construction into more complex membrane structures.

Notice that the construction presented here can be modified to simulate a non-deterministic TM by replacing the only two types of rules involving the transition function of the TM in a way to allow for a non-deterministic choice (due to having multiple rules in conflict):
\begin{align*}
  & \pevolvez{0}{q_{i,j,m+4,a}}{q_{i,j,m+5,(r,b,i+d)} b_{i+d,j,m+5}} & \text{for $q \in Q$,  $a \in \Sigma$,}\\
  & & \text{and $(r,b,i+d) \in \delta(q,a)$} \\
  & \pevolvez{0}{q_{i,j,m+5,(r,b,i+d)}}{r_{i+d,j+1,0}} & \text{for $q \in Q$ and $a \in \Sigma$}
\end{align*}
In the first rule the non-deterministic choice is remembered by writing it in the subscript. In this way, the only rule of the second kind that can fire is the one corresponding to the non-deterministic choice performed. We can then state the following theorem showing that a weaker uniformity condition is still sufficient to solve all $\NP$ problems with non-confluent systems:
\begin{theorem}
    $(\L,\L)$-uniform families of \emph{non-confluent} \emph{shallow} P~systems with active membranes with dissolution and without division can solve all problems in~$\NP$.
\end{theorem}

\section{Conclusions}
\label{sec:conclusions}

In this paper we showed that P~systems without charges can still solve any decision problem in the complexity class $\P$ even when the power of the Turing machines involved in the uniformity conditions is reduced. The TM simulation presented here is quite modular and can be embedded in more complex membrane structures. The resulting simulation is also efficient, requiring a slowdown of only a constant multiplicative factor.

However, some problems remain open, and the most prominent one is to study if the construction presented in~\cite{Leporati2014e} can be replicated for systems with charges, possibly adding an additional nesting level to accommodate for the different TM simulation technique. Such a result would show that even without charges the entire counting hierarchy can be computed in constant depth. This is another step in trying to understand what are the features that actually grant P~systems the power to go beyond the complexity class~$\P$ and, in some cases, beyond the entire polynomial hierarchy.

\bibliographystyle{splncs04}
\bibliography{Bibliography}

\end{document}